\documentclass[9pt,twocolumn,twoside]{forart}

\title{Systematic afterpulsing-estimation algorithms for gated avalanche photodiodes}

\author[1,*]{Carlos Wiechers}
\author[2]{Roberto Ram\'irez-Alarc\'on}
\author[2]{Oscar R. Mu\~niz-S\'anchez}
\author[1]{Pablo Daniel Y\'epiz}
\author[2,3]{Alejandro Arredondo-Santos}
\author[1]{Jorge G. Hirsch}
\author[1]{Alfred B U'Ren}

\affil[1]{Instituto de Ciencias Nucleares, Universidad Nacional Aut\'onoma de M\'exico, Apdo. Postal 70-543,  Cd. Mx. 04510 M\'exico.}
\affil[2]{Centro de Investigaciones en \'Optica, Loma del Bosque 115, Col. Lomas del Campestre, Leon 37150, Gto., Mexico}
\affil[3]{Departamento de F\'isica-DCI, Universidad de Guanajuato, P.O. Box E-143, 37150, Le\'on, Gto., M\'exico.}

\affil[*]{Corresponding author: carlos.wiechers@correo.nucleares.unam.mx}

\ociscodes{(040.1345) Avalanche photodiodes (APDs); (030.5260) Photon counting;(120.0120) Instrumentation, measurement, and metrology.}

\begin{abstract}
We present a method designed to efficiently extract optical signals from InGaAs avalanche photodiodes (APDs) operated in gated mode.  In particular, our method permits an estimation of the fraction of counts which actually results from the signal being measured, as opposed to being produced by noise mechanisms, specifically by afterpulsing. Our method in principle allows the use of InGaAs APDs at high detection efficiencies, with the full operation bandwidth, either with or without resorting to the application of a dead time. As we show below, our method can be used in configurations where afterpulsing exceeds the genuine signal by orders of magnitude, even near saturation.
The algorithms which we have developed are suitable to be used either in real-time processing of raw detection probabilities or in post-processing applications, after a  calibration step has been performed. The algorithms which we propose here can complement technologies designed for the reduction of afterpulsing.

\end{abstract}

\begin{document}

\maketitle

\section{Introduction} \label{Sec:1}

Single photon detectors based on avalanche photodiode (APD) technology can at present be operated at room-temperature with low power consumption and moderate to high efficiencies. They have been used widely in quantum key distribution (QKD) \cite{Lut:1,Sca:1}, light detection and ranging (LIDAR)\cite{Weh:1}, optical time-domain reflectometry (OTDR)\cite{Weg:1}, fiber optical sensing\cite{Mit:1}, biomedical applications\cite{Abo:1}, chemical sensing\cite{idQ:1}, photonics research\cite{RHad:1}, among other applications requiring weak optical signal sensing.

APDs are able to detect weak optical signals at the single-photon level. This is achieved using avalanche multiplication of photon-excited carriers, when they are  biased above their breakdown voltage. These avalanches (breakdown events) produce large enough currents that can be registered employing low-power electronic discriminators \cite{Yen:1}. APDs can be operated either in free-running mode or in gated mode\cite{Hai:1}. 

APDs have two main sources of noise\cite{Kang:1,Rib:1,His:1}: dark-count noise and afterpulsing noise. The first kind, dark-count noise, has its origin in thermal transitions and quantum inter-band tunnelling. Its probability of occurrence depends on the architecture and composition of the sensors, as well as on voltage and temperature settings. If these settings are kept constant, this noise is time-independent. If a significant number of carriers travel trough the APD during the activation gate (as in a breakdown event), they are able to populate carrier-traps in the APD multiplication layers. Later on, they can be spontaneously released, inducing spurious secondary breakdown events in the subsequent activation gates. This noise contribution is known as afterpulsing noise, which we will refer to simply as afterpulsing. In order to reduce the afterpulsing in the first generation of commercial InGaAs APDs, they should be gated using up to MHz frequencies, with the width of each activation gate set to a few ns. Furthermore, it is necessary to apply a dead-time of a few microseconds after each breakdown event, in order to allow de-trapping of carriers. This dead-time imposes a limit on the maximum count rate achievable in the APDs.  

Afterpulsing has strong implications on security of QKD systems: it limits the raw key generation rate, requiring the optimization of dead-time duration\cite{Stu:1,Lut:1}; it must be included in the quantum error correction of the distillation of secret keys\cite{Yos:1}, as well as in security proofs and tests \cite{Wie:1,Fer:1,Jai:1}.

Two important goals in the development of APD technology \cite{Itz:3,zha:3}  are to reduce afterpulsing, and to develop characterization techniques to improve the ability to discern genuine optical signal events from afterpulsing noise. A number of techniques have been proposed for the reduction of afterpulsing, based on: signal comparison\cite{Cho:1}, increasing the operating APD temperature\cite{Com:1,Kim:1}, passive quenching with an active reset\cite{Liu:1}, negative feedback\cite{Kor:2,Lun:1,Vin:1,Yan:1}, sub-Geiger avalanche gain operation\cite{Sti:2}, and photoionization of trapped carriers\cite{kra:1}. The most recent generation of single-photon detectors based on InGaAs APDs employs additional techniques to reduce afterpulsing, such as a self-differentiating post-processing~\cite{Hea:1,Yua:1} and sine-wave gating~\cite{Nam:1,Nam:2,Itz:3,Lu:1,Zha:2}. While these techniques lead to a reduction of afterpulsing, the full suppression of afterpulsing remains a challenge, in particular for higher detection efficiencies. Under intense background light the APD operation range is limited by the afterpulsing generated outside the activation gates \cite{Dal:1,Wie:1}. 

Alternative technologies for single-photon detection exist, such as: i) up-conversion, e.g. using periodically-poled LiNbO$_3$ (PPLN)\cite{Lan:1}, together with detection using Si APDs; and ii) superconducting nanowire detectors\cite{Gol:1}. 
They are also susceptible to afterpulsing\cite{Mar:1,Bur:1,Fuj:1}, as are other existing detection technologies such as  photomultiplier tubes \cite{Cam:1,Tor:1} and multi-pixel photon counters (MPPC)\cite{Du:1,oid:1}. 

Afterpulsing is a complex stochastic self-interacting phenomenon, which is proportional to the incoming light intensity. A number of techniques have been used to study it, including:  time interval analysis\cite{Yos:2,Fer:2,GHum:1}, double gate method\cite{Stu:1,Zha:1,Jen:1}, temporal distribution (background decay)\cite{Kor:1,Par:1,Dal:1}, in-gate effect of afterpulsing\cite{Hai:1,Yen:1,Hea:1}, corrections in $g^{(2)}(\tau)$ correlation measurements\cite{Kim:1,End:1,MZha:1,kra:1}, modified double-gate method (in order to study higher-order afterpulsing)\cite{Hai:1,Oh:1}, and other studies of  the dependence of afterpulsing on operation settings\cite{Kang:1,Com:1}.

Due to the electric field anisotropy in the  internal structure of APDs, there are ensembles of carrier-traps with associated energy distributions. In the multiple exponential decay function (MEDF) approach \cite{Kor:1,Stu:1,Itz:3,Fer:2,Hu:2} each time constant is related with a particular carrier trap energy\cite{Kor:1,Itz:2}. More sophisticated models introducing carrier-trap energy distributions are discussed in \cite{Itz:1,Kor:1,Hor:1}. The most widely used model to study afterpulsing is an effective single exponential decay function (SEDF) \cite{Kor:1,Stu:1,Itz:3,Fer:2,Hu:2,Itz:2,Hor:1}, with parameters determined by the corresponding mean values over the carrier-trap ensemble.

In what follows we use an SEDF model, which as we show below results in signal-extraction algorithms which can be implemented in real-time, thanks to its low time-processing cost. The formalism can be extended to MEDF or carrier-trap distribution models, as well, which are better suited for post-processing of acquired data. Having fully characterized the afterpulsing and dark-count noise parameters through a calibration procedure, the extraction of the photodetection signal probabilities is obtained with our algorithms from the raw mean detection probability measurements, even without the need to resort to the application of dead-times. The detection efficiency can be obtained with the use of an independently calibrated optical source. 

It is important  to point out that our algorithms presented in this paper cannot discriminate afterpulsing from the genuine photo-detection signal, on an event-by-event basis as required 
for sharing secrets keys in QKD systems.   Our algorithms allow a systematic estimation of the resepctive fractions of events due to genuine photo-detection, afterpulsing, and dark noise.   Our model provides a more complete picture of the physics behind avalanche photo-diodes, allowing the study of statistical and security issues of QKD systems based on this kind of detectors.  
 
This paper is organized as follows: in Section \ref{Sec:2}, we describe the afterpulsing models. The counting rule for raw probabilities without dead-time is presented in Section \ref{Sec:3}, and with dead-time in Section \ref{Sec:4}, using a time series analysis as an intermediate step. The experimental calibration procedure and the fitting algorithm are discussed in Section \ref{Sec:5}.  Also, we present an experimental application of our methodology in Section \ref{Sec:8}. Section \ref{Sec:9} summarizes our conclusions.  Finally, we include three appendices, in which we discuss special topics of this work: in appendix \ref{Ap:A} we describe the convergence in the afterpulsing probability introduced in \ref{Sec:3}, in appendix \ref{Ap:B} we present the calculations for fitting parameter errors and confidence intervals, and finally in appendix \ref{Ap:C} we show a phenomenological correction to the SEDF model, so as to account for an effect which we refer to as \emph{sub-counting}, induced by the electronic voltage discriminator, which converts the avalanche signal into a logic ON/OFF signal.

\section{Afterpulsing Models} \label{Sec:2}

We employ an SEDF model under the following assumptions:

\begin{itemize}
\item	Afterpulsing is cumulative in the absence of the application of dead-time.  Saturation of the electronic logic circuit in the APD modules occurs earlier, i.e. for a a lower incoming intensity, than saturation of the carrier-traps.  Thus, complete carrier-trap saturation is never reached.
\item	All breakdown events contribute on average equally to afterpulsing. This implies that the afterpulsing probability amplitudes are the same for all breakdown events.
\item	The de-trapping time constants and temporal probability amplitudes are the mean values over the corresponding carrier-trap ensembles.
\item	The activation gate duration is shorter than the de-trapping time constant, so that afterpulsing in each time gate is generated by breakdown events in previous gates.  Note that if this assumption is not valid, intra-gate afterpulsing corrections must be taken into account\cite{Hai:1,Yen:1,Hea:1}.  
\end{itemize}

\subsection*{Simple Afterpulsing model}

In the SEDF model the afterpulsing detection probability is modeled as

\begin{equation}
P_{af}(t)=\frac{Q}{\tau}\exp\left(-\frac{t}{\tau}\right), 
\label{eq:2:1}
\end{equation}

\noindent where $Q$ is a temporal constant related to the number of filled carrier-traps,  which determines the probability of generating an afterpulsing breakdown event \citep{Kor:1,Hu:2}.   $\tau$ is the de-trapping time parameter;  $Q/\tau\ (\le 1)$ represents the probability amplitude of producing an afterpulsing event.  The values of $Q$ and $\tau$ depend on the APD structure, and on the voltage and temperature settings.

As the APD operates in gated mode with a period $T$ and activation gate time-width $t_w$, it is convenient to write a discretized version of Eq. \ref{eq:2:1} as follows

\begin{equation}
P_{af}^{(n)}=\frac{Q}{\tau}\exp\left(-\frac{n}{F\tau}\right), 
\label{eq:2:2}
\end{equation}

\noindent where $F\equiv1/T$ is the operation frequency, and $P_{af}^{(n)}$ is the probability of occurrence of an afterpulsing avalanche, due to de-trapping of charges from an avalanche taking place at a moment in time corresponding to $n$ gates in the past.

\subsection*{Multiple-exponential decay functions}

In general, afterpulsing is generated by a distribution of carrier-traps with different time regimes, requiring an MEDF model. Since two different carrier-traps $A$ and $B$ can start avalanches independently, but both avalanches can occur at the same time gate, we use the probability addition rule 
\begin{equation}
P_A+P_B-P_AP_B=1-(1-P_A)(1-P_B).
\label{eq:2:3}
\end{equation}

For $K$ carrier-traps, with temporal constants  $Q_k$ and de-trapping time parameters  $\tau_k$, $k=1,...,K$, 
equation \ref{eq:2:2} is generalized as

\begin{equation}
P_{af}^{(n)}=1-\prod_{k=1}^K\left(1-\frac{Q_{k}}{\tau_{k}}\exp\left(-\frac{n}{F\tau_k}\right)\right),
\label{eq:2:4}
\end{equation}

The number of exponential decay functions needed to obtain a reliable description of the afterpulsing noise depends on the APD voltage and temperature settings\cite{Zha:1}. Nevertheless, in many situations it is worth using the simplest possible description of afterpulsing, with few parameters, in order to simplify the model and reduce the processing time. 

\section{Counting rule without dead-time} \label{Sec:3}

The total raw detection probability (click probability) includes genuine photodetection, dark counts, and afterpulsing. These are independent processes, whose probabilities are added according to the rule in Eq. \ref{eq:2:3}.

The initial gate of a finite sequence of gates has only two contributions, i.e. dark-count  noise ($P_{dc}$) and photodetection ($P_{ph}$), resulting in a click probability $P_c^{(1)}$ given by

\begin{equation}
P_c^{(1)}=1-(1-P_{dc})(1-P_{ph}). 
\label{eq:3:5}
\end{equation}

Photodetection of an attenuated laser beam is governed by the Poissonian statistics of a coherent state. For this kind of source, the photodetection probability is $P_{ph}=1-\exp(-\eta\mu)$; where, $\mu$ is the mean photon number per gate of the light source, and  $\eta$ is detector efficiency. 

The probability $P_c^{(1)}$ (photodetection and dark-count noise) remains constant throughout all gates in the sequence. We will thus regard $P_s \equiv P_c^{(1)}$ as a seed probability, which can cause afterpulsing noise in subsequent gates. 
At a given gate $n$, the click probability takes into account afterpulsing due to all breakdown events in previous gates, ending up with the following click probability

\begin{equation}
P_{c}^{(n)}=1-\left(1-P_{s}\right)\prod_{j=1}^{n-1}\left(1-P_{c}^{(n-j)}P_{af}^{(j)}\right).
\label{eq:3:6}
\end{equation}

We refer to the above expression of the click probability in the $n$th gate as the  \textit{forward building method} (FBM). The FBM may be used in order to characterize the parameters of the afterpulsing noise probability in sequences with a reduced number of gates.

If the number of gates in the sequence is sufficiently large, the raw count probability  $P_{c}^{(n)}$ converges to its asymptotic mean value $P_{c}^{(\infty)}$ at any activation gate. This convergence is discussed in Appendix \ref{Ap:A}. In this case we obtain the simplified expression 

\begin{equation}
P_{c}^{(\infty)}=1-\left(1-P_{s}\right)\prod_{j=1}^{\infty}\left(1-P_{c}^{(\infty)}P_{af}^{(j)}\right),
\label{eq:3:7}
\end{equation}
which is referred to as the \textit{backward building method} (BBM). 

The infinite product of equation \ref{eq:3:7} includes all afterpulsing contributions from previous gates; we call it the \textit{afterpulsing probability core} (APC). 

In the SEDF model, the APC resembles the \textit{q-Pochhammer} function\cite{q:1} $(a;q)_{\infty}$, defined as

\begin{equation}
\left(a;q\right)_{\infty}\equiv \prod_{j=0}^{\infty}\left(1-aq^j \right),
\label{eq:3:8}
\end{equation}

\noindent with $a \rightarrow P_{c}^{(\infty)}Q/\tau$ and $q\rightarrow \exp\left(-1/(F\tau)\right)$, except for the absence of the zeroth-order term. 

This $(a;q)_{\infty}$ function is widely used in $q$-analogue theory, in the description of exact statistical mechanics models\cite{Ren:1}, entropy of chaotic dynamics of many-particle systems\cite{Lig:1}, and avalanche-like processes in quantum networks\cite{Lar:1}. 

\section{Counting rule with dead-time} \label{Sec:4}

In most measurements employing InGaAs avalanche photodiodes a dead time $\Delta_t$ is set after each avalanche, so as to reduce the afterpulsing noise. In order 
to estimate the photodetection probability $P_{ph}$ from the raw detection probabilities $P_{c}^{(\infty)}$, the following considerations are employed:

\begin{enumerate}
\item After each detection event,  a dead time is applied. The first active gate which occurs once the dead time has expired is referred to as the 0th-gate. 

\item The probability of occurrence of an avalanche in the nth-gate due to afterpulsing originating from the detection event in question, without any cumulative correction, is

\begin{equation}
P^{(n)}_{af}=1-\prod_{k=1}^K\left(1-\frac{Q_{k}}{\tau_{k}}\exp\left(-\frac{n/F+\Delta_t}{\tau_k}\right)\right).
\label{eq:4:9}
\end{equation}

\item The probability of observing a raw detection event due to  photodetection, dark counts or afterpulsing, at the nth-gate, is
\begin{equation}
p_c^{(n)}=1-(1-P_s)(1-P^{(n)}_{af}),
\label{eq:4:10}
\end{equation}
for $n\ge 0$

\item At each detection event, the gate count is reset to zero.   The probability that the $n$th-gate is not inhibited by a dead time period is

\begin{equation}
p_g^{(n)}= \begin{cases}
	1  & \quad \text{if} \ \ n = 0 \\
    \prod_{j=0}^{(n-1)}\left(1-p_c^{(j)}\right) & \quad \text{if} \ \   n \ge 1.
  \end{cases}
\label{eq:4:11}  
\end{equation}

Note that if no dead-time is applied, $p_g^{(n)}=1$ for all values of $n$.

\item The asymptotic raw detection probability is obtained as an average over all probabilities

\begin{equation}
P_c^{(\infty)}=\frac{\sum_{n=0}^\infty p_c^{(n)}p_g^{(n)}}{\sum_{n=0}^\infty p_g^{(n)}}.
\label{eq:4:12}
\end{equation}
In this step we have performed a time interval analysis of afterpulsing, obtaining the time interval distribution.

\item We estimate the number of gates after the 0th-gate at which the subsequent detection event is expected, as 

\begin{equation}
N_g=\frac{\sum_{n=0}^\infty np_g^{(n)}}{\sum_{n=0}^\infty p_g^{(n)}}.
\label{eq:4:13}
\end{equation}

The time interval after the 0th-gate, at which a subsequent detection event is expected, is then $N_g T$ ($= N_g/F$).

\item Adding the dead time interval $\Delta_t$, which precedes the 0th-gate, we obtain the expected time between detection events
\begin{equation}
\Delta_{ht}=\Delta_t+N_g/F.
\label{eq:4:14}
\end{equation}

\item We now take into account possible afterpulsing noise originating not only from the last detection event but from previous detection events, so as to improve the counting rule. 
Including the information about the average time between detection events, Eq. \ref{eq:4:9} with cumulative afterpulsing correction becomes

\begin{equation}
P^{(n)}_{af}=1-\prod_{m=0}^M\prod_{k=1}^K\left(1-\frac{Q_{k}}{\tau_{k}}\exp\left(-\frac{g_{nm}}{\tau_k}\right)\right),
\label{eq:4:15}
\end{equation}

\noindent with $g_{nm}=nT+\Delta_t+m\Delta_{ht}$, and where $M$ is the maximum number of detection events which can originate cumulative afterpulsing noise. 

In cases where the dead time is sufficiently large, so as to suppress most  afterpulsing noise (i.e. $\Delta_{h t}> max\{\tau_k\} $), it is sufficient to restrict the analysis to 
$m=0$.

\end{enumerate}

\subsection*{Case: $\Delta_t >> max(\{\tau_k\})$}

If the dead time $\Delta_t $ is much longer than the decay times $\tau_k$, the afterpulsing noise becomes negligible. In this case $p_c^{(j)}$ becomes $P_s$, and  defining $p\equiv P_s$ and $q\equiv1-p$ the average number of gates between two subsequent detection events is expressed as

\begin{equation}
p_g^{(n)} = q^n. 
\label{eq:4:16}
\end{equation}
 
The sum over all  $ p_g^{(n)}$'s appearing in Eq.\ref{eq:4:13} becomes
 
\begin{eqnarray}
 \sum_{n=0}^\infty q^n & = & \frac{1}{p}, \label{eq:4:17}\\
 \sum_{n=0}^\infty nq^n & = & \frac{q}{p^2}, \label{eq:4:18}
\end{eqnarray}

\noindent and $N_g$ becomes

\begin{equation}
N_g=\frac{1-p}{p}. \label{eq:4:19}
\end{equation}

Employing Eq. \ref{eq:4:14}, the average time interval between subsequent detection events becomes
   
\begin{equation}
\Delta_{ht}=\frac{p(F\Delta_t-1)+1}{pF}.
\label{eq:4:20}
\end{equation}

Using equation \ref{eq:4:20}, we estimate the average number of detection events within a sampling time $T_S$ as

\begin{equation}
N_c = \frac{T_S}{\Delta_{ht}}  = \frac{pFT_S}{p(F\Delta_t-1)+1}. \label{eq:4:21}
\end{equation}

Note  that this expression remains valid in the limit $p\rightarrow 1$, giving $N_c^{max}=T_S/\Delta_t$. Note also that if $\Delta_t \le T$, there is effectively no dead-time because in that case, following any given detection event the next gate will always be available for detection.

We can invert the above equation, expressing the detection probability in terms of known quantities:  the sampling time $T_S$, the average number of detection counts $N_c$, the gating operation frequency $F$ and the dead-time $\Delta_t$

\begin{equation}
p=\frac{N_c}{T_S F-N_c(F\Delta_t-1)}.
\label{eq:4:22}
\end{equation}

Eq.  \ref{eq:4:22} is a generalization of the detection probability correction with dead-time given in reference \cite{RHad:1} for APDs in gated mode. Note that $T_S F$ is the total number of gates within the sampling time in the absence of dead time, and $N_c (F \Delta_t -1)$ is the total number of  gates removed during dead-time intervals. The difference in the denominator is the total number of active gates,  and represents the upper bound for the number of detection events. If the number of raw detection events  $N_c$ reaches this limit, there is a click at every gate, thus saturating the probability of detection ($p \rightarrow 1$).

\section{Afterpulsing characterization} \label{Sec:5}

In this section, we present the methodology that we have used in order to characterize dark-count noise, afterpulsing and the detection efficiency for a commercial InGaAs APD detector (id201 - IdQuantique).  Afterpulsing is an intrinsic phenomenon which appears for all APD module settings, but its effects are stronger for higher gating frequencies and higher efficiencies. In what follows we therefore concentrate our discussion on experiments carried out with higher detection efficiencies.  Considering that the detector parameters can change with detector age, calibration should be repeated perhaps a couple of times per year.

\subsection*{Calibration protocol}

\begin{figure}[h!]
\centering
\fbox{\includegraphics[width=0.95\linewidth]{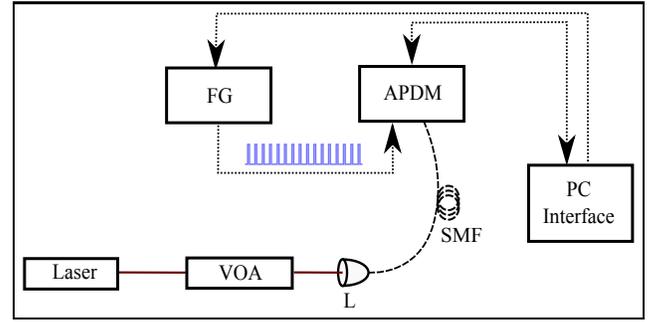}}
\caption{Schematic of the experimental setup for the afterpulsing characterization. VOA - variable optical attenuator, L - aspheric lens, SMF - single mode fiber, FG - function generator, APDM - APD module.}
\label{fig:1}
\end{figure}

The experimental setup used for calibration purposes is illustrated in Fig.  \ref{fig:1}. We have used as light source a continuous wave diode laser at 1550 nm (L: LDM1550 - Thorlabs), with its output power restricted by a calibrated variable optical attenuator (VOA\footnote{The VOA consists of a set of calibrated neutral filters for coarse power adjustment, together with the (calibrated) controlled separation between L and the fiber tip of SMF,  for fine power adjustment}). The attenuated laser beam is coupled, with the help of an aspheric lens (L), into a single mode fiber (SMF), which leads to the entrance port of the APD module (APDM)\cite{idQ:1}.  The APDM is externally triggered by  a function generator (FG), with a gate frequency up to $7.6$ MHz, thus bypassing the internal delay.  The FG and APDM are computer-controlled in our data acquisition routine.

In our experiments, described below, we selected the following APDM internal settings\citep{idQ:1}: \\
(a)  dead-time  $\Delta_t \in \{0,10\ \mu\mathrm{s}\}$,\\
 (b)  nominal gate temporal width $t_w = 2.5\ \mathrm{ns}$,\\ 
 (c) nominal efficiency $\eta \in \{0.20,\ 0.25\}$.\\

In addition, we have used the external parameters listed in Table \ref{tab:1} for the calibration, which correspond to 171 configurations. They consist of 19 different trigger frequency values ($F_i$), and two sets (S1 and S2) of 9 different mean photon number values, each,  per gate of the laser beam ($\mu_\nu$). S1 is used with $\eta=0.20$, and S2 is used with $\eta=0.25$. The mean photon number fluxes were determined taking into account the actual value of gate temporal width, $\sim0.8$ns, (as opposed to the nominal vale of $2.5$ns). These fluxes are reported in Table \ref{tab:1} and exhibit uncertainties of around 5\%.

\begin{table}[h!]
\begin{center}
\caption{\bf Operation parameters.}\label{tab:1}
\begin{tabular}{cc}
\hline
Parameter & Value(s) \\
\hline
 $\mu_\nu$ ($\times 10^{-2}$)  & S1: $\{0, 0.20, 0.72, 6.6, 14, 32, 67, 138,299 \}$    \\
 &  S2: $\{0, 0.20,	0.72, 5.5,	11.6, 25, 56, 107, 258 \}$  \\
$F_i$ (MHz) & $0.4$ to $7.6$, steps of $0.4$  \\
\hline
\end{tabular}
\end{center}
\end{table}

The average detection counts $\overline{N_{i\nu}}$, at a given operation frequency ($F_i$) and mean photon number ($\mu_\nu$),  are obtained through averaging over 120 data samples with a sampling time $T_S =1s$.  In the absence of dead-time, the experimental detection probabilities are calculated as

\begin{equation}
P_{c,i\nu}^{(e)}=\frac{\overline{N_{i\nu}}}{T_s F_i}.
\label{eq:5:23}
\end{equation}

When operating the APDM with dead time, we employ Eq.  \ref{eq:4:22}. The subindex $i$ in $P_{c,i\nu}^{(e)}$ refers to each gating frequency $F_i$, while the subindex $\nu$ refers to each mean photon number $\mu_{\nu}$.

\subsection*{Determining the afterpulsing parameters}

Our approach is to determine the model parameters, i.e. the probability amplitudes of afterpulsing detection $\{Q_k\}$ and de-trapping times $\{\tau_k\}$, from the calibration experimental runs.   For each flux value, the seed probability $P_{s,\nu}$  (see Eq. \ref{eq:3:5}) must also be determined. The dark count probability $P_{dc}$ is the seed probability when there is no photon flux.   

We determine these parameters by finding the parameter values which yield the best fit between the theoretical raw detection probabilities $P_{c,i\nu}^{(\infty)}$, across all frequency $F_i$ and mean photon number $\mu_{\nu}$ values, with the measured detection probabilities $P_{c,i\nu}^{(e)}$.  We have in practice restricted the number factors in APC, in such a manner that we include up to factor $j$  such that  $P_c^{(e)}P_{af}^{(j)}\geq\epsilon$, with $\epsilon = 10^{-10}$; we have verified that for this value of $\epsilon$ the output of our algorithm has already converged in all cases.

\begin{figure}[h!]
\centering
\includegraphics[width=0.90\linewidth]{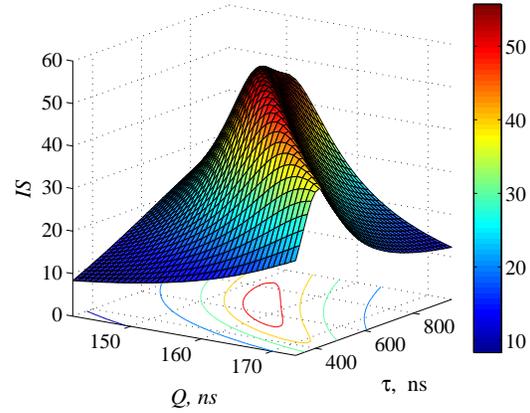}
\caption{The $IS$ function is used in order to obtain values for $Q$ and $\tau$. This example is related to results of Fig. \ref{fig:3}. 
}
\label{fig:2}
\end{figure}

The fitting algorithm maximizes the inverse of the sum of the squares of the relative deviations, over all the experimental measurements and their respective theoretical values, 

\begin{equation}
IS=\left[ \sum_{\nu=1}^{N_2}\sum_{i=1}^{N_1} \left( \frac{P_{c,i\nu}^{(e)}-P_{c,i\nu}^{(\infty)}}{P_{c,i\nu}^{(e)}} \right)^{2}   \right]^{-1}.
\label{eq:5:24}
\end{equation}
In the above expression, $N_2$ is the total number of different mean photon number values, and $N_1$ the number of different frequencies employed. 
This method is equivalent to the problem of minimizing the reciprocal quantity, $S_r^2=IS^{-1}$, however with certain important advantages:

\begin{enumerate}
\item It leads to a sharper peak (as compared with the corresponding trough for $S_r^2=IS^{-1}$) which aids the use of optimization algorithms (see Fig. \ref{fig:2}). 
  
\item The probability distribution of the previous point clarifies the possible presence of correlations among the carrier-trap parameters.   
\end{enumerate}

We have employed the following steps to maximize $IS$ function:

\begin{enumerate}
\item Use as initial seed probabilities the click probability value for the lowest trigger frequency $P_{s,\nu}=P_{c,1\nu}$.
\item Use as initial $Q$ value the inverse of the central frequency, $4$MHz, within the gating bandwidth ($0-8$MHz).
\item Use as initial $\tau$ twice the value of the original $Q$ value.
\item Find the $Q$ and $\tau$ values which jointly maximize the $IS$ function.
\item Find the optimal seed probabilities for each curve.
\item Repeat steps 4 and 5. Decreasing the parameter deviation, until its deviation is $<0.1\%$. 
\end{enumerate}

In our case, we have used a multi-dimensional maximum searching algorithm at each step. Each time the maximum value is reached, the grid size is reduced one order of magnitude and the grid spacing is reduced accordingly. 

From the function $S_r^2$ we can estimate the mean relative fitting error per degree of freedom,
\begin{equation}
\sigma_{f}=(rS^2/d.o.f)^{1/2}\times 100\%,
\label{eq:5:25}
\end{equation}

\noindent where   $d.o.f.=N_2\times N_1-2 K-N_2$. The number of different measurements is $N_2\times N_1$, and there are $2 K$ afterpulsing parameters $\{Q_{k},\tau_k\}$ and $N_2$ seed probabilities $\{P_{s,\nu}\}$.

The uncertainties in the estimation of the fitting parameters and their confidence intervals are obtained by analyzing the $S_r^2$ function, and using the first-order Taylor expansion of $P_c^{(\infty)}$ in the fitting parameters, as explained Appendix \ref{Ap:B}.

\section*{Results}
\subsection*{Nominal efficiency $\eta= 0.20$} 

\begin{figure}[h!]
\centering
\includegraphics[width=\linewidth]{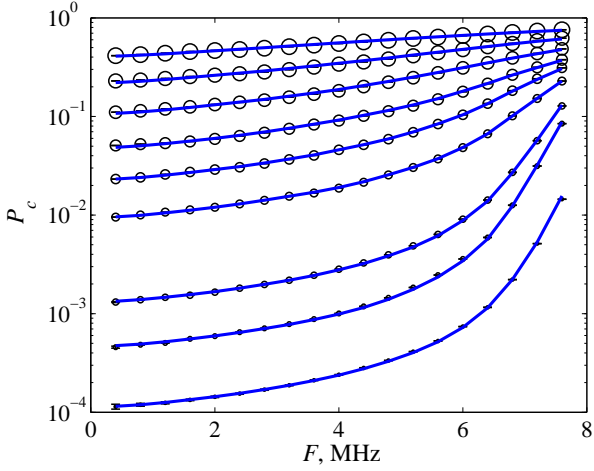}
\caption{Experimental data (black circles) and fitting curves (blue solid lines). Settings: $\eta=0.20$ and $t_w=2.5\ \mathrm{ns}$. Set: S1.}
\label{fig:3}
\end{figure}

The experimental data, black circles, and the corresponding fitting curves (solid blue lines) are presented in Fig. \ref{fig:3}; note that the circle radius indicates the seed probability, with a larger radius corresponding to a larger probability.    Through the optimization of the $IS$ function over the $(Q,\tau)$ plane (see Fig. \ref{fig:2}), we have found that there is a region where these parameters yield a good fit with the calibration experimental runs; note that in this case there is a significant correlation between the $Q$ and $\tau$ parameters.  We use the maximum value of the $IS$ function to find the best-fit parameters.  The optimization for each seed probability ($P_{s,\nu}=P_{s}(\mu_\nu)$) was performed separately on the corresponding curve. 

The fitting parameters, dark count probabilities and afterpulsing parameters, are presented in Table \ref{tab:2}. Their respective t-test values (mean divided by standard deviation) to be compared with the threshold value $t_c=2.85$ for the t-student distribution with $160$ degrees of freedom and confidence level of $99.5\%$ is shown in the last  column.  Note that all parameters exhibit a t-test value which comfortably fulfills the condition $t > t_c$.

The effective APDM efficiency, $\eta_r$, is obtained from the photodetection probabilities and their respective previously-determined flux values for each curve (see table \ref{tab:1}). Once $P_{s,\nu}$ is determined, using equations \ref{eq:3:5} and \ref{eq:3:7}, we obtain $P_{ph,\nu}$. The mean efficiency for set S1 is $\eta_r = 0.169 \pm 0.010$, to be compared with the nominal efficiency selected in the APD, i.e. $\eta= 0.20$.

\begin{table}[h!]
\centering
\caption{\bf Fitting parameters. $t_w=2.5\ \mathrm{ns}$ $\eta= 0.20$. Set: S1}
\begin{tabular}{ccc}
\hline
Parameter & Values & t-test ($t>2.85$) \\
\hline
$P_{dc}$ & $(1.144\pm 0.072)\ \times 10^{-4}$ & $15.8 $\\
$Q$  & $(157.6\pm 1.0)$  ns & $150.5 $ \\ 
$\tau$ & $(637.8\pm 25.8)$  ns & $24.7 $\\
$\eta_r$ & $ 0.169 \pm 0.010$ & $16.9$\\
$\sigma_{f}$ & $1.86\%$ &  \\
\hline
\end{tabular}
  \label{tab:2}
\end{table}

\begin{figure}[h!]
\centering
\includegraphics[width=0.90\linewidth]{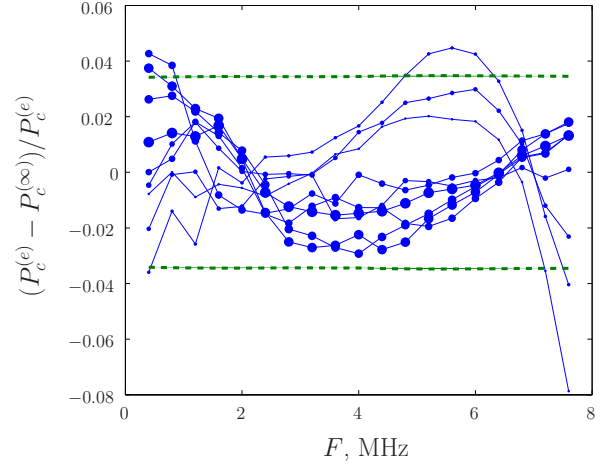}
\caption{Relative deviations (blue dot and solid lines) and confidence intervals (green dashed lines), corresponding to the results of Fig. \ref{fig:3}.}
\label{fig:4}
\end{figure}

Substituting  the parameter values shown in Table \ref{tab:2} in Eq. \ref{eq:3:7}, the fitting curves (blue solid lines) are compared with the measured detection probabilities (black circles)  shown in Fig. \ref{fig:3}, plotted as a function of the gating frequency $F$. As the photon flux range covers four orders of magnitude, the scale employed for the detection probability $P_c$ is logarithmic; note the remarkable agreement between the model and the experimental results. To asses in more detail the quality of the theoretical description, the relative deviation between each experimental data point and its modeled value is presented in Fig. \ref{fig:4} (blue dot/solid-line); note that the dot size as in Fig. \ref{fig:3} indicates the seed probability, with a larger dot corresponding to a larger probability.   We have also  included the mean confidence intervals (over the $\nu$ index) for each frequency (green dashed line). The quality of the fit is in all cases excellent, with deviations of 4\% at the most, and with a low value of the fitting error $\sigma_f=1.86\%$ (see Eq. \ref{eq:5:25}).

We have performed the F-test for each curve with 19 configurations and 3 fitting parameters each. The critical value for a 99.5\% confidence level is $F_c=7.51$. The F-test values for each curve are $22.3,\ 80.0,\ 231,\ 1.31\times 10^4,\ 306,\ 296,\ 137,\ 37.9,\ 38.7$. Clearly all of these values pass the test $(>F_c)$ comfortably. 

\subsection*{Nominal efficiency $\eta= 0.25$} 

\begin{figure}[h!]
\centering
\includegraphics[width=\linewidth]{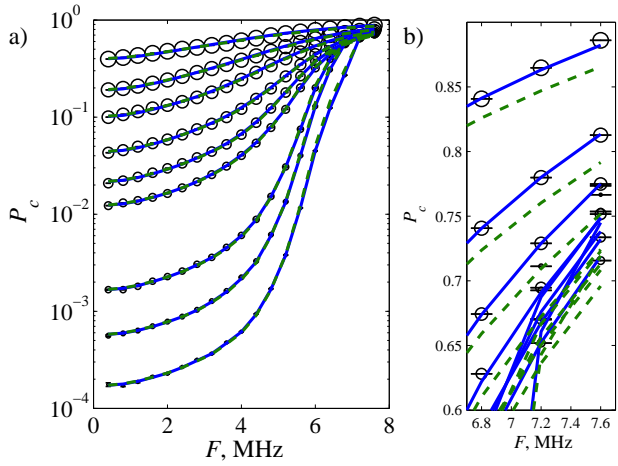}
\caption{a) Experimental data (black circles), fit with correction (blue solid lines) and fit without correction (green dashed lines). Settings: $t_w= 2.5\ ns$ and $\eta=0.25$ Set: S2.  b) Close up of panel a, for large gating frequencies}  
\label{fig:5}
\end{figure}

We have carried out  our APDM characterization for a nominal efficiency of $\eta= 0.25$, in addition to the efficiency of $\eta= 0.20$ presented above.  In this case, we have used the second set of seed probability values  S2, with a nominal gate temporal width of  $t_w= 2.5\ \mathrm{ns}$; our results are presented in Fig. \ref{fig:5}a. As in the previous case, in the logarithmic scale the fit is excellent for all the photon flux values. At this efficiency, high gate frequencies (above 6 MHz) approach saturation making it challenging, but still possible, to distinguish different seed probabilities (see Fig. \ref{fig:5}b).  The fitting parameters are presented in Table \ref{tab:3}. 

\begin{table}[h!]
\centering
\caption{\bf Fitting parameters. $t_w=2.5\ \mathrm{ns}$ $\eta= 0.25$. Set: S2.}
\begin{tabular}{ccc}
\hline
\multicolumn{3}{l}{Without correction.}\\
\hline
Parameter & Values  & t-test ($t>2.85$) \\
\hline
$P_{dc}$ & $(1.706\pm 0.459) \ \times 10^{-4}$ & $3.7 $\\
$Q_{1}$ & $(230.0\pm 2.2)$ ns & $106.9 $ \\ 
$\tau_{1}$ & $(464.7\pm 17.6)$ ns & $26.4 $\\
$\eta_r\ $ & $ 0.199 \pm 0.017$ & $11.7$\\
$\sigma_{f}$ & $3.12\%$ &  \\
\hline
\end{tabular}
  \label{tab:3}
\end{table}

\begin{figure}[h!]
\centering
\includegraphics[width=0.9\linewidth]{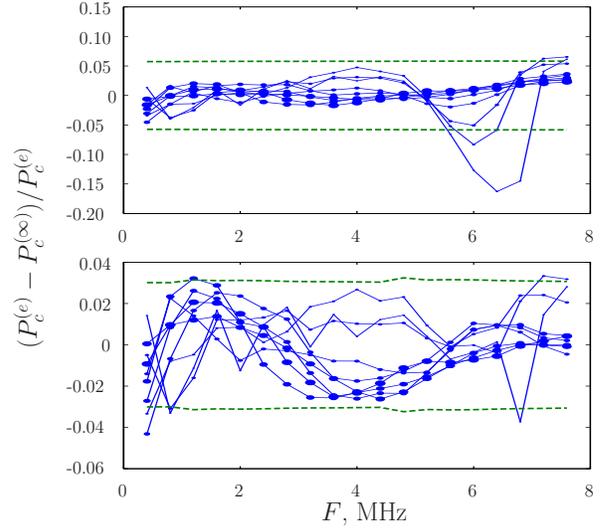}
\caption{Deviations (blue dot and solid lines) and confidence intervals (green dashed lines) corresponding to Fig. \ref{fig:5}. Upper graph: experimental data vs model without sub-counting effect. Bottom graph: experimental data vs model with sub-counting effect.}
\label{fig:6}
\end{figure}

The relative deviations between the experimental data and the curves derived from the model, using the best-fit parameters,   are shown in Fig. \ref{fig:6} (top). The F-test for the various curves leads to the following values:  $21.7,\ 25.2,\ 37.4,\ 163,\ 103,\ 74.0\ 86.8,\ 70.5,\ 37.6$. All these values pass the test $(>F_c)$ comfortably, with $F_c=7.51$.

At higher gate frequencies and lower incident fluxes the deviations are larger than  5\%, leading the model to overestimate the corresponding experimental measurements; we refer to this as a sub-counting effect. A phenomenological model of the de-trapping time parameter can be constructed to correct for the sub-counting effect, as presented in Appendix \ref{Ap:C}, and employed to improve the fit, as shown in Fig. \ref{fig:6} (bottom). The F-test parameter for these curves has the following values, with 19 configurations and 6 fitting parameters ($F_c=5.79$):  $525,\ 333,\ 601,\ 7.75\times 10^3,\ 5.08\times 10^3,\ 2.40\times 10^3,\ 1.34\times 10^3,\ 618,\ 166$;  clearly, explicitly incorporating the sub-counting effect leads to a major improvement in the quality of the fits.   To rule out that this sub-counting effect is a property of one specific detector, we have verified that it occurs in three different APD modules of the same model (id201 from idQuantique\citep{idQ:1}).

\subsection*{Signal-to-noise ratio} 

We estimate the total noise probability with the expression

\begin{equation}
P_{n,i\nu}=1-(1-P_{dc})\prod_{j=1}^{\infty}\left(1-P_{c,i\nu}^{(e)}P_{af}^{(j)}\right),
\label{eq:5:26}
\end{equation}
\noindent which  is estimated by adding the dark counts probability to the total afterpulsing contribution. Once the afterpulsing parameters are determined, the probability of detecting dark counts $P_{dc}$ is obtained from the seed probability with mean photon flux $\mu=0$ (so that only dark counts can act as seed); this corresponds to the lowest curves in Figs. \ref{fig:3} and \ref{fig:5}.


As can be observed in Figs.  \ref{fig:3} and \ref{fig:5}, for each mean photon flux the detection probability $P_c$ exhibits a smooth dependence on the gate frequency  $F$, which is well reproduced by the model. Considering that afterpulsing noise is not random but strongly correlated in time with any previous signals, the afterpulsing noise can essentially map the intensity optical signal distribution to a similar afterpulsing noise distribution. This in fact makes afterpulsing a particularly difficult source of noise to deal with in practice, since at first sight it can easily be mistaken for genuine signal. This implies the need for an algorithm such as the one presented in this paper for the correct estimation of the fraction of counts which can be attributed to genuine detection events.

Combining the expressions for the probability of detection $P_c$, Eq. \ref{eq:3:7}, the seed probability $P_s$, Eq. \ref{eq:3:5}, and the noise probability $P_n$, Eq. \ref{eq:5:26}, the photodetection probability can be expressed as 

\begin{equation}
P_{ph}=1-\frac{1-P_{c}}{1-P_{n}}. \label{eq:5:27}
\end{equation}

For each photon flux value $\mu_{\nu}$, there are $N_1$ different gate frequencies $F_i$, with measured detection probabilities  $P_{c,i\nu}^{(e)}$, associated with the same photodetection probability  $P_{ph,\nu}$.  To take into account the fluctuations of the measured values along the fitted curve, we estimate the photodetection probability as an average over all the frequencies along the same curve

\begin{equation}
P_{ph,\nu}=1-\frac{1}{N_1}\sum_{i=1}^{N_1}\frac{1-P_{c,i\nu}^{(e)}}{1-P_{n,i\nu}}.
\label{eq:5:28}
\end{equation}
In order to calculate the product in Eq. \ref{eq:5:26}, we have in practice restricted the number factors in such a manner that we include only factors $j$ which fulfil  $P_c^{(e)}P_{af}^{(j)}\geq\epsilon$, with $\epsilon = 10^{-10}$.

\begin{figure}[h!]
\centering
\includegraphics[width=0.90\linewidth]{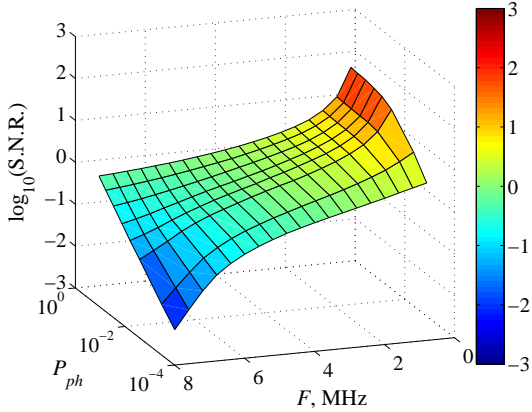}
\caption{Signal-to-noise ratio in logarithmic scale, derived from Fig. \ref{fig:3} after applying our noise discrimination algorithm.}
\label{fig:7}
\end{figure}

Employing the values obtained for the photodetection probability, Eq. \ref{eq:5:28} , and the noise probability, Eq. \ref{eq:5:26}, we evaluate the signal-to-noise ratio 
(SNR) as

\begin{equation}
\mathrm{SNR} _{i\nu}=\frac{P_{ph,\nu}}{P_{n,i\nu}}.
\label{eq:5:29}
\end{equation}

Fig. \ref{fig:7} shows the logarithm base 10 of the SNR. In the regions of low photodetection probability and high gate frequency the SNR is very small, with afterpulsing noise dominating over photodetections by more than to two orders of magnitude.

\subsection*{Comparing models}

In what follows, we compare our own model, to be referred to as Model 1, with other existing models.   Most afterpulsing models in the literature are able to describe the afterpulsing probability in situations where it is smaller than the photodetection probability.    Model 2 (ref. \cite{GHum:1}) does not incorporate cumulative afterpulsing and is used in time-series analysis;  our methodology of Section \ref{Sec:4}  reduces to Model 2 if the cumulative effect is disregarded. Model 3 (ref. \cite{Oh:1}) includes cumulative effects, but is unable to describe the saturation due to afterpulsing. 

\begin{figure}[h!]
\centering
\includegraphics[width=0.90\linewidth]{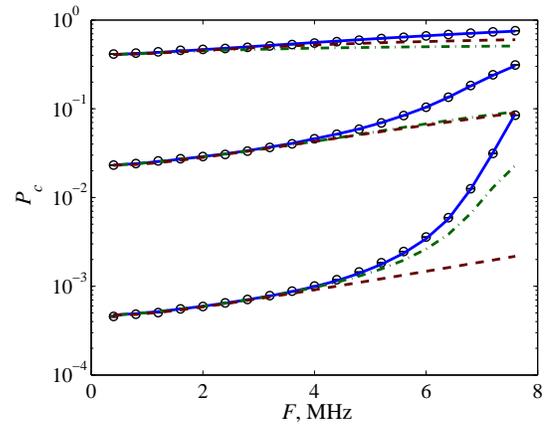}
\caption{Comparison between experimental data (black circles) and afterpulsing models: Models  1 (solid blue line), Model 2 (red dashed line) and Model 3 (green dash-dot line), for three different seed probabilities}
\label{fig:8}
\end{figure}

In Fig. \ref{fig:8}, we present a comparison of our own model (Model 1) with Models 2 and 3, for three different sets of measured probabilities, selected from Fig. \ref{fig:3}. For Models 2 and 3, the best fitting parameters allow for a very good description of configurations  with gating frequencies up to 4 MHz. In their respective frequency ranges, the three models converge to statistically the same seed-probabilities. This  is the case for small gate frequencies, for which the afterpulsing probability is smaller than photodetection probability. When the gate frequency exceeds  4 MHz these three models make different predictions, with Model 1 clearly showing a much better agreement with our experimental data over the full operation bandwidth, as compared to Models 2 and 3. Model 2 has the optimal afterpulsing parameters $Q=(209.8\pm 4.5)\ \mathrm{ns}$ and $\tau=(372\pm 20) \ \mathrm{ns}$; and for Model 3 the afterpulsing parameters are $Q=(173.3\pm 1.9)\ \mathrm{ns}$ and $\tau=(459.3\pm 39.5) \ \mathrm{ns}$.

\subsection*{Results with dead time}

\begin{figure}[h]
\centering
\includegraphics[width=0.90\linewidth]{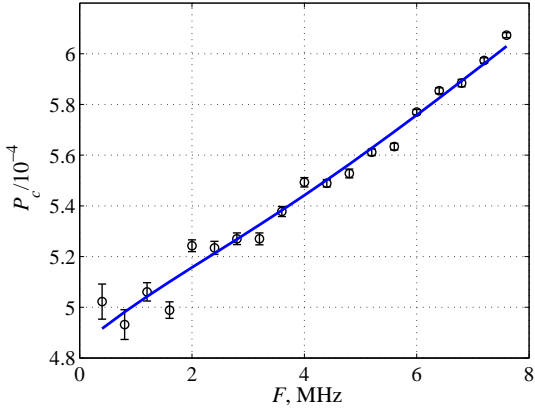}
\caption{Detection probabilities vs. gate frequency, with dead time. 
Experimental data (black circles), fit (solid blue line). Settings: $t_w= 2.5\ \mathrm{ns}$ , $\eta=0.20$ and $\Delta_t=10\  \mu s$ }
\label{fig:9}
\end{figure}

The afterpulsing behavior with dead-time in the APDM can be evaluated with the same setup as described in Fig. \ref{fig:1}, by selecting a dead-time of $\Delta_t$ in the APDM settings. Imposing a  dead-time after each detection event in the APDM reduces the afterpulsing contribution, at the expense of a reduction of the maximum count rate. As we can observe in Fig. \ref{fig:9}, there is a remanent afterpulsing contribution, which at the highest gating frequencies represents around 20\% of the total signal. The fitting is performed using the model described in Section \ref{Sec:4} with cumulative afterpulsing.

The afterpulsing parameter values obtained with this methodology are $Q=(38.57 \pm 1.10)\ \mathrm{ns}$ and $\tau=(201 \pm 185) \ \mu \mathrm{s}$. The $\tau$ parameter exhibits a large dispersion, which means that the remaining ensemble of carrier-traps  has a wide energy distribution. Using the efficiency and dark count probability of Table \ref{tab:2}, we obtain $\mu=(2.19\pm 0.13)\times 10^{-3}$ photons per gate.  This configuration is less sensitive to afterpulsing contributions of carrier-traps with small de-trapping times; they essentially do not contribute to the  mean values of the SEDF parameters.

\section{Application example} \label{Sec:8}

As a proof-of-principle demonstration of this methodology, we have analyzed signals from a photon pair source based on the spontaneous parametric down conversion (SPDC) process, in a type-I configuration. We have chosen the operation settings of the APDM as: gating frequency $F=6.0$ MHz, nominal gate temporal width $t_w=2.5$ns and nominal detection efficiency $\eta=0.20$.  In this particular example, the afterpulsing probability is greater than the photodetection probability, and there is no observable sub-counting effect. This means that we can use the SEDF model without sub-counting correction in the BBM.

\begin{figure}[h!]
\centering
\includegraphics[width=0.95\linewidth]{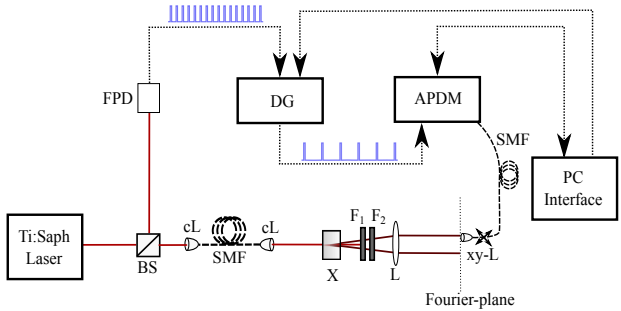}
\caption{Characterization of SPDC process. Schematic setup. BS - Beamsplitter; FPD - Fast photodiode; DG - Delay Generator; APDM - APD Module; SMF - Single mode fiber; L -Lens; cL - fiber-coupling lens; $xy$-L - $xy$-scanning fiber-coupling lens; F$_1$ - Low-Pass Filter; F$_2$ - Band-Pass Filter; X - $\beta$-BBO crystal.}
\label{fig:10}
\end{figure}

\begin{figure*}[!ht]
\centering
\includegraphics[trim = 0mm 30mm 0mm 30mm, width=0.9\linewidth]{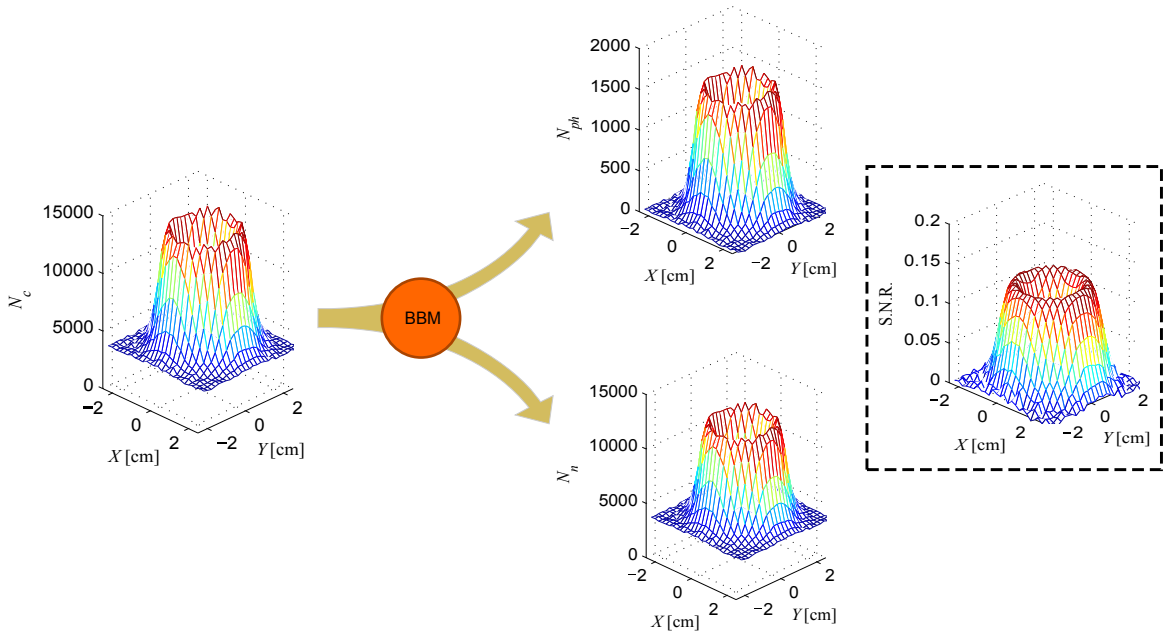}
\caption{Filtering results: Using the BBM, photodetection counts $N_{ph}$ are discriminated from noise counts $N_n$, when the total counts are measured $N_c$ . The SNR distribution (on $XY$-plane) is presented inside the dashed square. Settings:$\eta=0.20$, $t_w=2.5\ \mathrm{ns}$, $F=6.0\ \mathrm{MHz}$. }
\label{fig:11}
\end{figure*}

\subsection*{Data acquisition for the SPDC photon-pair source without dead time}

The SPDC photon-pair source used is shown in  Fig. \ref{fig:10}.  A $\beta$-barium borate ($\beta$-BBO) crystal (X) is pumped by a femtosecond Ti-sapphire laser beam, centered at $775$nm so as to produce frequency-degenerate telecommunications-band photon pairs (at $1550$nm).  The laser output is spatially filtered so that the pump beam can be described to good approximation as a Gaussian beam.  Under these conditions,  the SPDC light exhibits the well-known type-I annular spatial distribution.

In order to suppress the pump, the signal and idler photon pairs are transmitted through a low-pass filter, (F$_1$) transmitting wavelengths $\lambda>980$ nm;  and through a $1550\pm 5\ \mathrm{nm}$ band-pass filter (F$_2$).   A lens (L), with focal length  $f=10\ \mathrm{cm}$ placed at a distance $f$ from the crystal, defines a Fourier plane at a further  distance $f$ from the lens.  The full SPDC transverse spatial intensity distribution was recorded by means of a fiber tip which scans the Fourier plane with the help of a 2-dimensional precision motor, leading to the APDM \cite{Ram:1}.  The electronic signal used as external trigger for the APDM (which defines the gating frequency) was obtained as follows: i) we employed the electronic pulse train produced by a fast photodiode sensing a portion of the laser beam, and ii) we used a pre-scaler/delay circuit (DG: DG645 - Stanford Research Systems) which selects one out of every 15 electronic pulses so as to reduce the  (electronic rather than optical) repetition rate from $90\ \mathrm{MHz}$ to $6\ \mathrm{MHz}$.   At this gating frequency, the APDM exhibits a considerable afterpulsing contribution (which in fact exceeds the optical signal). Since the internal delay circuit of the APDM settings is bypassed\cite{idQ:1}  the internal delay is fixed, allowing APDM operation using its full bandwidth. Therefore the pre-scaler/delay circuit permits the synchronization of the SPDC optical signals with the APD gates.

In our experiment, we collected 10 samples of spatially-resolved counts taken over a matrix of transverse positions defined by a $2.2\ \mathrm{cm}\times 2.2\ \mathrm{cm}$ window with $0.2\ \mathrm{cm}$ steps.  The experimental detection probabilities $P_c^{(e)}$ are obtained using Eq.  \ref{eq:5:23}. 

\subsection*{Complete noise discrimination results}

One of the key results of this paper is Eq. \ref{eq:3:7}, which is rather convenient in order to perform the noise discrimination (subtraction), once we have fully characterized the afterpulsing and dark noise contributions (see Section \ref{Sec:5}). This permits the extraction of the optical detection probability $P_{ph}$ from the raw total detection probability$P_c^{(\infty)}$, which is replaced with $P_c^{(e)}$ in Eq. \ref{eq:3:7}, 

\begin{equation}
P_{ph}=1-\frac{\left(1-P_{c}^{(e)}\right)}{\left(1-P_{dc}\right)\prod_{j=1}^{\infty}\left(1-P_{c}^{(e)}P_{af}^{(j)}\right)},
\label{eq:6:30}
\end{equation}

\noindent which is a very simple and powerful, analytic expression. With its help, we  extract the genuine photo-detection signals, i.e. removing noise contributions from the overall measured detector output. 

The initial task is to calculate the APC, using the afterpulsing and dark noise parameters of Table \ref{tab:2}, together with the gating frequency of 6.0 MHz. In order to calculate the product in Eq. \ref{eq:6:30}, we have in practice restricted the number of factors in such a manner that we include all factors $j$  such that  $P_c^{(e)}P_{af}^{(j)}\geq\epsilon$, with $\epsilon = 10^{-10}$.

Substituting the parameter values shown in Table \ref{tab:2} in Eq.  \ref{eq:6:30}, we can separate the photodetection probability $P_{ph}$ from all noise contributions, including  afterpulsing and dark noise. We depict in Fig. \ref{fig:11} an illustration of this process, where the number of counts ($N_x$) is given by the respective probability $P_x$ multiplied by the gating frequency $F$. In this plot the subindex $x=c$ denotes total counts, $x=ph$ denotes photodetection counts, and $x=n$ denotes noise counts.  Inside the dashed square in Fig. \ref{fig:11}, we show the SNR distribution corresponding to the SPDC ring, using Eq. \ref{eq:5:29}. 

\section{Conclusions} \label{Sec:9}

We have presented a noise discrimination methodology for avalanche photo diode modules (APDM) operated in gated mode, designed to quantify the relative weights of the optical, dark count, and afterpulsing contributions to the total number of detection events.   This methodology, which is simple and robust,  is applicable to the operation of APDMs without, as well as with, dead-time (Sections \ref{Sec:3} and Section \ref{Sec:4}, respectively).  Our methodology is furthermore comparatively less resource-intensive than other noise discrimination techniques and permits the estimation of both dark count and afterpulsing noise components over the entire operation bandwidth.  In our approach we can  estimate the full set of APDM parameters, including detection efficiency,  using a simple calibration experimental procedure. 

Our method allows us to reliably operate an APDM and extract the actual optical signal  at all gating frequencies allowed, and with any detection efficiency setting, as long as the detector does not fully saturate.  The application of our method renders commercial APDMs considerably more noise tolerant and permits the extraction of the actual optical signal from the total counts even in situations when afterpulsing noise dominates.  

Our paper also contributes to a greater understanding of the physics behind afterpulsing. In contrast to dark count noise, which is random, the afterpulsing noise essentially maps the spatial or temporal optical signal distribution to a similar afterpulsing noise distribution.  We have demonstrated the use of our methodology in the context of an experiment for the determination of the spatial transverse count distribution of a spontaneous parametric down conversion photon-pair source.   The ability to correctly identify the fraction of counts due to genuine photodetection, as made possible by our method, implies that the experimenter can take advantage of the greater genuine signal which results from larger gating frequencies and detector efficiencies, unhampered by the adverse effects of afterpulsing.

\begin{appendix}
\section{APPENDIX: Convergence Limit of APC}\label{Ap:A}

In this appendix we discuss the convergence of $P_c^{(n)}$, see Eq.  \ref{eq:3:6}, to the value $P_c^{(\infty)}$.  We expect $P_c^{(n)}$ to converge to a certain value because the afterpulsing contribution from past detection events decreases with the time interval since their occurrence.  

In our calculations, we use as cut-off for the infinite product in Eq. \ref{eq:3:6},  conserving  factors such that  $P_c^{(n-j)}P_{af}^{(j)} \geq \epsilon$, with $\epsilon=10^{-10}$.   

\begin{figure}[h]
\centering
\includegraphics[width=0.95\linewidth]{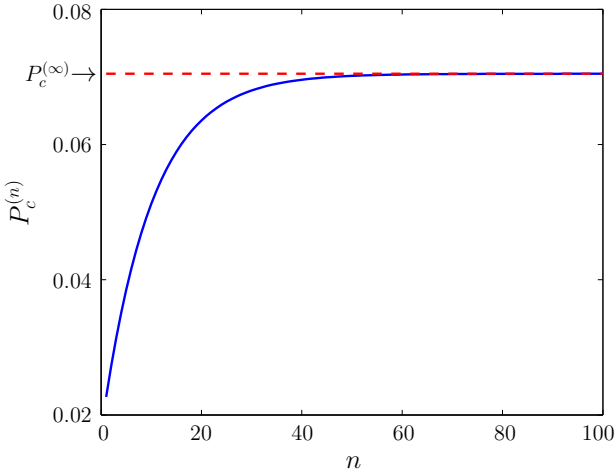}
\caption{Convergence of $P_c^{(n)}$ using FBM.}
\label{fig:12}
\end{figure}

The convergence behavior of  $P_c^{(n)}$ is illustrated in Fig. \ref{fig:12} for the specific case of a  flux of $\mu=0.14$ photons per gate, with the following choice of parameters: detection nominal efficiency $\eta=0.20$, nominal gate temporal width $t_w=2.5\ \mathrm{ns}$, and gating frequency $F=5.2$ MHz.

Let us now analyze the convergence of the APC product, in Eq.  \ref{eq:3:7}.  To that end, let us  consider the logarithm of the APC, $W$,   

\begin{equation}
W=\ln\left(\prod_{j=1}^{\infty}\left(1-P_{c}^{(\infty)}P_{af}^{(j)}\right)\right)=\sum_{j=1}^{\infty}\ln\left(1-P_{c}^{(\infty)}P_{af}^{(j)}\right).
\label{eq:A:31}
\end{equation}

Note that because $P_{c}^{(\infty)} \leq 1$, $P_{af}^{(j)}  \leq 1$, and $P_{af}^{(j+1)}<P_{af}^{(j)}$, the ratio between two consecutive terms  in Eq. \ref{eq:A:31} is less than unity (in particular, as $j\rightarrow \infty$), i.e.

\begin{equation}
\lim_{j \rightarrow \infty} \frac{\ln\left(1-P_{c}^{(\infty)}P_{af}^{(j+1)}\right)}{\ln\left(1-P_{c}^{(\infty)}P_{af}^{(j)}\right)} < 1.
\label{eq:A:32}
\end{equation}

This guarantees that $W$ converges, and that the APC product converges as well.

\section{APPENDIX: Error parameter estimation}\label{Ap:B}

In this appendix, we show how to estimate the fitting parameter errors. We regard $P_c^{(\infty)}$ as a multi-dimensional function. We define a multi-dimensional  vector of parameters

\begin{equation}
\mathbf{r}  \equiv ( \{P_{S,\nu}\},\{Q_k \},\{\tau_k \},\cdots ),
\label{eq:B:33}
\end{equation} 

\noindent and we can write the first-order Taylor expansion of the click probability as

\begin{equation}
P_c^{(\infty)}(\mathbf{r})=P_c^{(\infty)}(\mathbf{r}_0)+\mathbf{\nabla}_{\mathbf{r}} P_c^{(\infty)}(\mathbf{r}_0)\cdot\Delta \mathbf{r}.
\label{eq:B:34}
\end{equation}

Inserting Eq. \ref{eq:B:34} in the definition of $S_r^2$, the inverse of $IS$ in Eq. \ref{eq:5:24}, we obtain
\begin{equation}
S_r^2=\sum_{\nu=1}^{N_2}\sum_{i=1}^{N_1}\left( \frac{P_{c,i\nu}^{(e)}-P_{c,i\nu}^{(\infty)}(\mathbf{r}_0)-\mathbf{\nabla}_{\mathbf{r}} P_{c,i\nu}^{(\infty)}(\mathbf{r}_0) \cdot\Delta \mathbf{r}}{P_{c,i\nu}^{(e)}} \right)^2.
\label{eq:B:35}
\end{equation}

In order to simplify the notation, we rewrite the last Eq. as

\begin{equation}
S_r^2=\sum_{\nu=1}^{N_2}\sum_{i=1}^{N_1}\sum_{u=1}^{U}\sum_{v=1}^{U}\left( y^{i\nu}-X_u^{i\nu}\beta^u \right)\left( y_{i\nu}-X_{i\nu}^v\beta_v \right).
\label{eq:B:36}
\end{equation}
where we have use the following notation

\begin{eqnarray}
\mathbf{y}=y_{i\nu} &  = & \frac{P_{c,i\nu}^{(e)}-P_{c,i\nu}^{(\infty)}(\mathbf{r}_0)}{P_{c,i\nu}^{(e)}}, \label{eq:B:37}\\
\mathbf{X}=X_{i\nu s} & = & \frac{\partial_{\mathbf{r},s} P_{c,i\nu}^{(\infty)}(\mathbf{r}_0)}{P_{c,i\nu}^{(e)}},\label{eq:B:38} \\
\boldsymbol\beta=\beta_s & = & \Delta r_s,\label{eq:B:39} 
\end{eqnarray}
with $s=\{u,v\}$.  Using the  least squares methodology with matrix notation, we obtain

\begin{equation}
\mathbf{C}=C_u^v  =  \left[\sum_{\nu=1}^{N_2}\sum_{i=1}^{N_1}X_u^{i\nu}X_{i\nu}^v \right]^{-1}, \label{eq:B:40}
\end{equation}
which is used to find the solutions
\begin{equation}
\beta_s  = \sum_{v=1}^{U} C_s^v  \left[\sum_{\nu=1}^{N_2}\sum_{i=1}^{N_1}X_v^{i\nu}y_{i\nu}\right], \label{eq:B:40b}
\end{equation}

We can estimate by iteration, the solution as follow,

\begin{equation}
r_{(k+1),s}  = r_{(k),s}(1+\beta_s). \label{eq:B:40c}
\end{equation}

Also, Eq. \ref{eq:B:40} is used to evaluate the relative error in the parameters,
\begin{equation}
\delta \beta_s =\sigma_f\sqrt{C_s^s},\label{eq:B:41}
\end{equation}
where $C_s^s$ is the diagonal of $C_u^v$.

For each parameter in $\mathbf{r}$, we calculate its propagation error as
\begin{equation}
\delta r_s = r_s\delta \beta_s.\label{eq:B:42}
\end{equation}

The average confidence interval in the frequency domain is expressed as
\begin{equation}
\mathbf{G}=G_{i\nu}  =  \sum_{u=1}^{U}X_u^{i\nu}C_u^uX_{i\nu}^u , \label{eq:B:43}
\end{equation} 
which gives
\begin{equation}
\delta y_{i\nu}=\sigma_f(1+\sqrt{G_{i\nu}}) . \label{eq:B:44}
\end{equation}

\section{APPENDIX: Sub-counting effect}\label{Ap:C}

We have observed that there is a region of gating frequencies for which our model over estimates the observed click probability.  This effect, which we refer to as sub-counting,  is apparent for small seed probabilities while it becomes negligible for a sufficiently large seed probability.    The relative deviations between experimental data and the fitted functions are depicted in Fig. \ref{fig:6} (top), where  the sub-counting effect can be appreciated. In order to model this sub-counting dependence, we propose a Gaussian function dependence for the de-trapping time parameter,

\begin{equation}
\tau_k=\tau_{uk}+\alpha_{k}(1-P_c^{(e)})\exp\left(-\frac{(f-f_{a,k})^2}{2f_{b,k}^2}\right),
\label{equ:C:45}
\end{equation}

where $f_{a,k}$ is the  centroid of the Gaussian, $f_{b,k}$ is the standard deviation, and $\alpha_k$ is the time amplitude correction. There is also a phenomenological dependence on the experimental click probability.

\begin{table}[h!]
\centering
\caption{\bf Fitting parameters. $t_w=2.5\ \mathrm{ns}$ $\eta= 0.25$. Set: S2.}
\begin{tabular}{ccc}
\hline
\multicolumn{3}{l}{With correction.}\\
\hline
Parameter & Values  & t-test ($t>2.85$) \\
\hline
$P_{dc}$ & $(1.730\pm 0.080)\ \times 10^{-4} $ & $21.8 $\\
$Q$ & $(250.3\pm 1.7)$ ns& $150.6 $ \\ 
$\tau_{u}$ & $(401.5\pm 7.2)$ ns & $55.5 $\\
$\alpha$ & $(293.6\pm 28.9)$ ns & $10.2 $ \\
$f_{a}$ & $(7.188\pm 0.275)$ MHz & $26.2 $ \\
$f_{b}$ & $(1.230\pm 0.158)$ MHz& $7.8 $ \\
$\eta_r\ $ & $ 0.198 \pm 0.017$ & $11.6$\\
$\sigma_{f}$ & $1.62\%$ &  \\
\hline
\end{tabular}
  \label{tab:4}
\end{table}

Using Eq. \ref{equ:C:45}, the relative deviation between experimental points and fitted data is reduced, as can be seen in Fig. \ref{fig:6} (bottom). This leads to a
better description of the afterpulsing behavior under more general conditions. The fitting results are presented in Table \ref{tab:4}, in this case the threshold level for the t-student distribution is $t_c=2.85$ with $d.o.f.=157$ and confidence level of 99.5\%. Also, the fitting error $\sigma_f$ is reduced from 3.12\%, without correction, to 1.62\%, with correction. 

This sub-counting effect is also observed for set S1 with a detection efficiency of $\eta=0.20$ for small seed probabilities (see Fig. \ref{fig:4}), in a range of frequencies shifted to larger values as compared to S2 (see Fig. \ref{fig:6}).

\end{appendix}

\bigskip
\noindent This work was supported by SEP-CONACyT and RedTC-CONACyT, M\'{e}xico, and PAPIIT (UNAM) grant number IN105915.

\bibliography{Wiechers_afterpulsing}

\begin{thebibliography}{10}
\newcommand{\enquote}[1]{``#1''}

\bibitem{Lut:1}
N.~L\"{u}tkenhaus and A.~J. Shields, \enquote{Focus on quantum cryptography:
  Theory and practice,} New J. Phys. \textbf{11}, 045005 (2009).

\bibitem{Sca:1}
V.~Scarani, H.~Bechmann-Pasquinucci, N.~J. Cerf, M.~Du\v{s}ek,
  N.~L\"{u}tkenhaus, and M.~Peev, \enquote{The security of practical quantum
  key distribution,} Rev. of Mod. Phys. \textbf{81}, 1301 (2009).

\bibitem{Weh:1}
A.~Wehr and U.~Lohr, \enquote{Airborne laser scanning -- an introduction and
  overview,} ISPRS J. Photogramm. \textbf{54}, 68--82 (1999).

\bibitem{Weg:1}
M.~Wegmuller, F.~Scholder, and N.~Gisin, \enquote{Photon-counting otdr for
  local birefrigence and fault analysis in the metro environment,} Journal of
  Lightwave Technology \textbf{22}, 1--11 (2004).

\bibitem{Mit:1}
M.~W. Mitchell, J.~S. Lundeen, and A.~M. Steinberg, \enquote{Super-resolving
  phase measurements with a multiphoton entangled state,} Nature \textbf{429},
  161--164 (2004).

\bibitem{Abo:1}
A.~F. Abouraddy, M.~B. Nasr, B.~E.~A. Saleh, A.~V. Sergienko, and M.~C. Teich,
  \enquote{Quantum-optical coherence tomography with dispersion cancellation,}
  Phys. Rev. A \textbf{65}, 053817 (2002).

\bibitem{idQ:1}
M.~Desert, \emph{id201 datasheet}, idQuantique, v.4.1 ed. (2009).

\bibitem{RHad:1}
R.~H. Hadfield, \enquote{Single-photon detectors for optical quantum
  information applications,} Nature Photonics \textbf{3}, 696--705 (2009).

\bibitem{Yen:1}
H.~T. Yen, S.~D. Lin, and C.~M. Tsai, \enquote{A simple method to characterize
  the afterpulsing effect in single photon avalanche photodiode,} Journal of
  Applied Physics \textbf{104}, 054504 (2008).

\bibitem{Hai:1}
M.~Hai-Qiang, Y.~Jian-Hui, W.~Ke-Jin, L.~Rui-Xue, and Z.~Wu,
  \enquote{Afterpulsing characteristics of ingaas/inp single photon avalanche
  diodes,} Chinese Physics B \textbf{23}, 120308 (2014).

\bibitem{Kang:1}
Y.~Kang, H.~X. Lu, and Y.-H. Lo, \enquote{Dark count probability and quantum
  efficiency of avalanche photodiodes for single-photon detection,} Applied
  Physics Letters \textbf{83}, 2955--2957 (2003).

\bibitem{Rib:1}
G.~Ribordy, J.-D. Gautier, H.~Zbinden, and N.~Gisin, \enquote{Performance of
  ingaas/inp avalanche photodiodes as gated-mode photon counters,} Applied
  Optics \textbf{37}, 2272--2277 (1998).

\bibitem{His:1}
P.~A. Hiskett, G.~S. Buller, A.~Y. Loudon, J.~M. Smith, I.~Gontijo, A.~C.
  Walker, P.~D. Townsend, and M.~J. Robertson, \enquote{Performance and design
  of ingaas/inp photodiodes for single-photon counting at 1.55 $\mu$m,} Applied
  Optics \textbf{39}, 6818--6829 (2000).

\bibitem{Stu:1}
D.~Stucki, G.~Ribordy, A.~Stefanov, H.~Zbinden, J.~G. Rarity, and T.~Wall,
  \enquote{Photon counting for quantum key distribution with peltier cooled
  ingaas/inp apds,} Journal of Modern Optics \textbf{48}, 1967--1981 (2001).

\bibitem{Yos:1}
A.~Yoshizawa, R.~Kaji, and H.~Tsuchida, \enquote{After-pulse-discarding in
  single-photon detection to reduce bit errors in quantum key distribution,}
  Optics Express \textbf{11}, 1303--1309 (2003).

\bibitem{Wie:1}
C.~Wiechers, L.~Lydersen, C.~Wittmann, D.~Elser, J.~Skaar, C.~Marquardt,
  V.~Makarov, and G.~Leuchs, \enquote{After-gate attack on a quantum
  cryptosystem,} New Journal of Physics \textbf{13}, 013043 (2011).

\bibitem{Fer:1}
T.~F. da~Silva, G.~B. Xavier, G.~P. Tempor$\tilde{\mathrm{a}}$o, and J.~P.
  von~der Weid, \enquote{Real-time monitoring of single-photon detectors
  against eavesdropping in quantum key distribution systems,} Optics Express
  \textbf{20}, 18911--18924 (2012).

\bibitem{Jai:1}
N.~Jain, E.~Anisimova, I.~Khan, V.~Makarov, C.~Marquardt, , and G.~Leuchs,
  \enquote{Trojan-horse attacks threaten the security of practical quantum
  cryptography,} New J. Phys. \textbf{16}, 123030 (2014).

\bibitem{Itz:3}
M.~A. Itzler, X.~Jiang, M.~Entwistle, K.~Slomkowski, A.~Tosi, F.~Acerbi,
  F.~Zappa, and S.~Cova, \enquote{Advances in ingaasp-based avalanche diode
  single photon detectors,} Journal of Modern Optics \textbf{58}, 174--200
  (2011).

\bibitem{zha:3}
J.~Zhang, M.~A. Itzler, H.~Zbinden, and J.-W. Pan, \enquote{Advances in
  ingaas/inp single-photon detector systems for quantum communication,} Light:
  Science \& Applications \textbf{4}, e286 (2015).

\bibitem{Cho:1}
S.-B. Cho and S.-K. Kang, \enquote{Weak avalanche discrimination for gated-mode
  single-photon avalanche photodiodes,} Optics Express \textbf{19},
  18510--18515 (2011).

\bibitem{Com:1}
L.~C. Comandar, B.~Fr\"{o}hlich, J.~F. Dynes, A.~W. Sharpe, M.~Lucamarini,
  Z.~L. Yuan, R.~V. Penty, and A.~J. Shields, \enquote{Gigahertz-gated
  ingaas/inp single-photon detector with detection efficiency exceeding 55$\%$
  at 1550?nm,} J. Appl. Phys. \textbf{117}, 083109 (2015).

\bibitem{Kim:1}
Y.-S. Kim, Y.-C. Jeong, S.~Sauge, V.~Makarov, and Y.-H. Kim, \enquote{Ultra-low
  noise single-photon detector based on si avalanche photodiode,} Review of
  Scientific Instruments \textbf{82}, 093110 (2011).

\bibitem{Liu:1}
M.~Liu, C.~Hu, J.~C. Campbell, Z.~Pan, and M.~M. Tashima, \enquote{Reduce
  afterpulsing of single photon avalanche diodes using passive quenching with
  active reset,} IEEE JOURNAL OF QUANTUM ELECTRONICS \textbf{44}, 430--434
  (2008).

\bibitem{Kor:2}
SPIE, \emph{Low temperature performance of free-running InGaAs/InP singlephoton
  negative feedback avalanche diodes}, vol. 9114 of \emph{Advanced Photon
  Counting Techniques VIII} (SPIE, 2014). Boris Korzh and Hugo Zbinden.

\bibitem{Lun:1}
T.~Lunghia, C.~Barreiroa, O.~Guinnarda, R.~Houlmanna, X.~Jiangb, M.~A. Itzlerb,
  and H.~Zbindena, \enquote{Free-running single-photon detection based on a
  negative feedback ingaas apd,} Journal of Modern Optics \textbf{59},
  1481--1488 (2012).

\bibitem{Vin:1}
\emph{Probability Distribution and Noise Factor of Solid State Photomultiplier
  Signals with Cross-Talk and Afterpulsing}, 2009 IEEE Nuclear Science
  Symposium Conference Record (NSS/MIC) (IEEE, 2009). S. Vinogradov and T.
  Vinogradova and V. Shubin and D. Shushakov and K. Sitarsky.

\bibitem{Yan:1}
Z.~Yan, D.~R. Hamel, A.~K. Heinrichs, X.~Jiang, M.~A. Itzler, and T.~Jennewein,
  \enquote{An ultra low noise telecom wavelength free running single photon
  detector using negative feedback avalanche diode,} Rev. Sci. Instrum.
  \textbf{83}, 073105 (2012).

\bibitem{Sti:2}
M.~Stip\v{c}evi\'{c}, \enquote{Commercially available geiger mode single-photon
  avalanche photodiode with a very low afterpulsing probability,}
  arXiv:1505.04407v1 [physics.ins-det]  (2015).

\bibitem{kra:1}
\emph{Photoionization of Trapped Carriers in Avalanche Photodiodes to Reduce
  Afterpulsing During Geiger-Mode Photon Counting}, no. CMGG4 in Conference on
  Lasers and Electro-Optics/Quantum Electronics and Laser Science and Photonic
  Applications Systems Technologies, Technical Digest (CD) (OSA, 2005). Michael
  A. Krainak.

\bibitem{Hea:1}
C.~Healey, I.~Lucio-Martinez, M.~R.~E. Lamont, X.~Mo, and W.~Tittel,
  \enquote{Characterization of an ingaas/inp single-photon detector at 200 mhz
  gate rate,} arXiv:1105.3760v1 [quant-ph]  (2011).

\bibitem{Yua:1}
Z.~L. Yuan, B.~E. Kardynal, A.~W. Sharpe, and A.~J. Shields, \enquote{High
  speed single photon detection in the near infrared,} Applied Physics Letters
  \textbf{91}, 041114 (2007).

\bibitem{Nam:1}
N.~Namekata, S.~Sasamori, and S.~Inoue, \enquote{800 mhz single-photon
  detection at 1550-nm using an ingaas/inp avalanche photodiode operated with a
  sine wave gating,} Optics Express \textbf{14}, 10043--10049 (2006).

\bibitem{Nam:2}
N.~Namekata, S.~Adachi, and S.~Inoue, \enquote{1.5 ghz single-photon detection
  at telecommunication wavelengths using sinusoidally gated ingaas/inp
  avalanche photodiode,} Optics Express \textbf{17}, 6275--6282 (2009).

\bibitem{Lu:1}
Z.~Lu, W.~Sun, Q.~Zhou, J.~Campbell, X.~Jiang, and M.~A. Itzler,
  \enquote{Improved sinusoidal gating with balanced ingaas/inp single photon
  avalanche diodes,} Optics Express \textbf{21}, 16716--16721 (2013).

\bibitem{Zha:2}
SPIE, \emph{2.23 GHz gating InGaAs/InP single-photon avalanche diode for
  quantum key distribution}, vol. 7681 of \emph{Advanced Photon Counting
  Techniques IV}. Jun Zhang and Patrick Eraerds and Nino Walenta and Claudio
  Barreiro and Rob Thew and Hugo Zbinden.

\bibitem{Dal:1}
A.~D. Mora, D.~Contini, A.~Pifferi, R.~Cubeddu, A.~Tosi, and F.~Zappa,
  \enquote{Afterpulse-like noise limits dynamic range in time-gated
  applications of thin-junction silicon single-photon avalanche diode,} Applied
  Physics Letters \textbf{100}, 241111 (2012).

\bibitem{Lan:1}
C.~Langrock, E.~Diamanti, R.~V. Roussev, Y.~Yamamoto, M.~M. Fejer, and
  H.~Takesue, \enquote{Highly efficient singlephoton detection at communication
  wavelengths by use of upconversion in reverse-proton-exchanged periodically
  poled $linbo_3$ waveguides,} Optics Letters \textbf{30}, 1725 -- 1727 (2005).

\bibitem{Gol:1}
G.~N. Gol'tsman, O.~Okunev, G.~Chulkova, A.~Lipatov, A.~Semenov, K.~Smirnov,
  B.~Voronov, A.~Dzardanov, C.~Williams, and R.~Sobolewski, \enquote{Picosecond
  superconducting single-photon optical detector,} Appl. Phys. Lett.
  \textbf{79}, 705 (2001).

\bibitem{Mar:1}
F.~Marsili, F.~Najafi, E.~Dauler, R.~J. Molnar, and K.~K. Berggren,
  \enquote{Afterpulsing and instability in superconducting nanowire avalanche
  photodetectors,} Applied Physics Letters \textbf{100}, 112601 (2012).

\bibitem{Bur:1}
V.~Burenkov, H.~Xu, B.~Qi, R.~H. Hadfield, and H.-K. Lo,
  \enquote{Investigations of afterpulsing and detection efficiency recovery in
  superconducting nanowire single-photon detectors,} Journal of Applied Physics
  \textbf{113}, 213102 (2013).

\bibitem{Fuj:1}
M.~Fujiwara, A.~Tanaka, S.~Takahashi, K.~Yoshino, Y.~Nambu, A.~Tajima, S.~Miki,
  T.~Yamashita, Z.~Wang, A.~Tomita, , and M.~Sasaki, \enquote{Afterpulse-like
  phenomenon of superconducting single photon detector in high speed quantum
  key distribution system,} Optics Express \textbf{19}, 19562--19571 (2011).

\bibitem{Cam:1}
L.~Campbell, \enquote{Afterpulse measurement and correction,} Review of
  Scientific Instruments \textbf{63}, 5794 (1992).

\bibitem{Tor:1}
S.~Torre, T.~Antonioli, and P.~Benetti, \enquote{Study of afterpulse effects in
  photomultipliers,} Review of Scientific Instruments \textbf{54}, 1777 (1983).

\bibitem{Du:1}
Y.~Du and F.~Reti\`{e}re, \enquote{After-pulsing and cross-talk in multi-pixel
  photon counters,} Nuclear Instruments and Methods in Physics Research A
  \textbf{596}, 396--401 (2008).

\bibitem{oid:1}
\emph{Study of afterpulsing of MPPC with waveform analysis}, International
  workshop on new photon-detectors (PD07) (Proceedings of Science, 2007). H.
  Oide and H. Otono and S. Yamashit and T. Yoshiok and H. Hano and T. Suehiro.

\bibitem{Yos:2}
A.~Yoshizawa, R.~Kaji, and H.~Tsuchida, \enquote{Quantum efficiency evaluation
  method for gated-mode single-photon detector,} Electronics Letters
  \textbf{38}, 1468--1469 (2002).

\bibitem{Fer:2}
T.~F. da~Silva, G.~B. Xavier, , and J.~P. von~der Weid, \enquote{Real-time
  characterization of gated-mode single-photon detectors,} Quantum Electronics,
  IEEE Journal of \textbf{47}, 1251 -- 1256 (2011).

\bibitem{GHum:1}
G.~Humer, M.~Peev, C.~Schaeff, S.~Ramelow, M.~Stip\v{c}evi\'{c}, and R.~Ursin,
  \enquote{A simple and robust method for characterization of afterpulsing in
  single photon detectors,} J. Lightwave Technol. \textbf{33}, 3098--3107
  (2015).

\bibitem{Zha:1}
J.~Zhang, R.~Thew, J.-D. Gautier, N.~Gisin, and H.~Zbinden,
  \enquote{Comprehensive characterization of ingaas-inp avalanche photodiodes
  at 1550 nm with an active quenching asic,} IEEE Journal of Quantum
  Electronics \textbf{45}, 792--799 (2009).

\bibitem{Jen:1}
K.~E. Jensen, P.~I. Hopman, E.~K. Duerr, E.~A. Dauler, J.~P. Donnelly, S.~H.
  Groves, L.~J. Mahoney, K.~A. McIntosh, K.~M. Molvar, A.~Napoleone, D.~C.
  Oakley, S.~Verghese, C.~J. Vineis, and R.~D. Younger, \enquote{Afterpulsing
  in geiger-mode avalanche photodiodes for 1.06 $\mu$m wavelength,} Applied
  Physics Letters \textbf{88}, 133503 (2006).

\bibitem{Kor:1}
B.~Korzh, T.~Lunghi, K.~Kuzmenko, G.~Boso, and H.~Zbinden,
  \enquote{Afterpulsing studies of low noise ingaas/inp single-photon negative
  feedback avalanche diodes,} arXiv:1411.0653 [physics.ins-det]  (2014).

\bibitem{Par:1}
A.~Para, \enquote{Afterpulsing in silicon photomultipliers: Impact on the
  photodetectors characterization,} arXiv:1503.01525 [physics.ins-det]  (2015).

\bibitem{End:1}
J.~Enderlein and I.~Gregor, \enquote{Using fluorescence lifetime for
  discriminating detector afterpulsing in fluorescence-correlation
  spectroscopy,} Review of Scientific Instruments \textbf{76}, 033102 (2005).

\bibitem{MZha:1}
M.~Zhao, L.~Jin, B.~Chen, Y.~Ding, H.~Ma, and D.~Chen, \enquote{Afterpulsing
  and its correction in fluorescence correlation spectroscopy experiments,}
  Applied Optics \textbf{42}, 4031--4036 (2003).

\bibitem{Oh:1}
J.~Oh, C.~Antonelli, M.~Tur, and M.~Brodsky, \enquote{Method for characterizing
  single photon detectors in saturation regime by cw laser,} Optics Express
  \textbf{18}, 5906--5911 (2010).

\bibitem{Hu:2}
C.~Hu, X.~Zheng, J.~C. Campbell, B.~M. Onat, X.~Jiang, and M.~A. Itzler,
  \enquote{Characterization of an ingaas/inp-based single-photon avalanche
  diode with gated-passive quenching with active reset circuit,} Journal of
  Modern Optics \textbf{58}, 201--209 (2011).

\bibitem{Itz:2}
M.~A. Itzler, R.~Ben-Michael, K.~S. C.~F.~Hsu, S.~C. A.~Tosi, F.~Zappa, and
  R.~Ispasoiu, \enquote{Single photon avalanche diodes (spads) for 1.5 $\mu$m
  photon counting applications,} Journal of Modern Optics \textbf{54}, 283--304
  (2007).

\bibitem{Itz:1}
M.~A. Itzler, X.~Jiang, and M.~Entwistle, \enquote{Power law temporal
  dependence of ingaas/inp spad afterpulsing,} Journal of Modern Optics
  \textbf{59}, 1475--1480 (2012).

\bibitem{Hor:1}
D.~B. Horoshko, V.~N. Chizhevsky, and S.~Y. Kilin, \enquote{Full-response
  characterization of afterpulsing in single-photon detectors,} arXiv:1409.6752
  [quant-ph]  (2014).

\bibitem{q:1}
G.~E. Andrews, \emph{\textit{q}-Series: Their Development and Application in
  Analysis, Number Theory, Combinatorics, Physics and Computer Algebra}, 66
  (CBMS Regional Conference Series in Mathematics, 1986).

\bibitem{Ren:1}
E.~J.~J. van Rensburg, \emph{The Statistical Mechanics of Interacting Walks,
  Polygons, Animals and Vesicles} (Oxford, 2015), 2nd ed.

\bibitem{Lig:1}
T.~M. Liggett, \emph{Interacting Particle Systems}, vol. 276 of
  \emph{Grundlehren der mathematischen Wissenschaften} (Springer, 2005), 1st
  ed.

\bibitem{Lar:1}
D.~B. Larremore, M.~Y. Carpenter, E.~Ott, and J.~G. Restrepo,
  \enquote{Statistical properties of avalanches in networks,} Physical Review E
  \textbf{85}, 066131 (2012).

\bibitem{Ram:1}
R.~Ram\'{i}rez-Alarc\'{o}n, H.~Cruz-Ram\'{i}rez, and A.~B. U'Ren,
  \enquote{Effects of crystal length on the angular spectrum of spontaneous
  parametric downconversion photon pairs,} Laser Phys. \textbf{23}, 055224
  (2013).

\end{thebibliography}

\end{document}